\documentclass[12pt]{article}          
\usepackage{graphicx,epsf,subfigure,pstricks,pst-node,psfrag,amsthm,amssymb,amsmath,url,inputenc}
\numberwithin{equation}{section}

\begin{document}

\title{\textbf{The Efficient Shrinkage Path: Maximum Likelihood of Minimum MSE Risk}}
\author{Robert L. Obenchain \\ Risk Benefit Statistics}
\date{February 2024}
\maketitle

\begin{abstract}
\noindent A new \textit{generalized} ridge regression shrinkage path is proposed
that is as short as possible under the restriction that it must pass through the vector of
regression coefficient estimators that make the overall Optimal Variance-Bias Trade-Off under
Normal distribution-theory. Five distinct types of ridge \textit{TRACE} diagnostics plus other
graphics related to efficient estimates are motivated and illustrated here. These visualizations
provide invaluable data-analytic insights and improved self-confidence to researchers and data
scientists fitting linear models to ill-conditioned (confounded) data.

\noindent {\it Keywords: Linear Models, Generalized Ridge Regression, Normal-theory Maximum
Likelihood, Mean-Squared-Error Risk, TRACE diagnostic displays, R-functions.}
\end{abstract}

\section{Regularization Perspective} 
Regularization of linear models in machine learning is focused here on how
\textit{generalized} ridge \textit{TRACE} displays enable researchers to literally ``see''
the effects of variance-bias trade-offs on measures of mean-squared-error (MSE) risk that
accompany all changes in estimated $\beta-$coefficients. Normal-theory Maximum Likelihood
(ML) estimates exist for linear models when the number, $p$, of non-constant $x-$predictor
variables is strictly less than the number of observations, $n$. The $X-$predictor matrix is
then said to be ``\textit{narrow}'', and the ordinary least-squares (OLS) estimator is uniquely
determined. The additional assumption that at least one OLS residual, $y_i-\hat{y}_i$, is
non-zero assures that the estimated error-variance, $\hat{\sigma}^2$, is strictly positive.

Gunst (2000, p. 62) wrote: ``Although ridge regression is widely used in the application
of regression methods today, it remains as controversial as when it was first introduced.'' Many
early proposals for choosing the ``k-factor'' defining the $1-$parameter shrinkage path of Hoerl
and Kennard (1970a,b) were indeed based mainly on \textit{heuristics}. In sharp contrast, shrinkage
based upon Normal-theory Maximum Likelihood estimation has a firm theoretical foundation,
Obenchain (1975, 1977, 1978, 2022).

The \textit{generalized} ridge estimators of ${\beta}-$coefficients that are most-likely
to have minimum MSE risk under Normal-theory are of the known form discussed here in $\S 5$; see
Thompson (1968). No GRR shrinkage path with fewer than $p-$parameters will always pass through
this ``best'' point-estimate. A new \textit{Efficient} $p-$parameter shrinkage-path in the form
of a two-piece linear spline with a single interior Knot at the ``best'' point-estimate
is proposed in $\S 6$ and illustrated graphically in the following sections, $\S 7-\S 9$.
 
Ridge \textit{TRACE} displays for this new efficient path and other visualizations are
implemented in version $2.3$ of the \textit{RXshrink} R-package, Obenchain (2023).
Each TRACE typically displays estimates of $p \geq 2$ quantities that change as
\textit{shrinkage} occurs. The ``coef'' TRACE displays all $p$ fitted linear-model
$\beta-$coefficients, and the ``rmse'' TRACE plots \textit{relative} mean-squared-error
estimates given by the diagonal elements of the MSE-matrix divided by the OLS-estimate
of $\sigma^2$. Whenever shrinkage becomes excessive, the ``infd'' TRACE displays estimated
direction-cosines for the \textit{inferior-direction} in $X-$space along which shrunken
coefficients have higher MSE Risk than OLS estimates, Obenchain (1978). The two remaining types
of \textit{TRACE diagnostics}, named ``spat'' and ``exev'', refer to the $p$ rotated axes defining
the \textit{principal-coordinates} of the given (confounded) $x-$predictors. All estimates
displayed in \textit{TRACEs} are ML or unbiased or have correct-range when $p < n-3$.

\section{Shrinkage Estimation} 

Linear models and the OLS estimator, $\hat{\beta}_0$, of $\beta-$coefficients can be placed in
a canonical form that is easy to generalize when defining the shrinkage-estimators
of interest here. We assume that the $y-$outcome vector has been both centered and
re-scaled to have an observed mean of zero and variance 1, that each column
of the $X-$matrix ($n \times p$) has been standardized in this same way, and that the
resulting $X-$matrix has full (column) \textit{rank} $= p$ that is $\geq 2$ and $\leq (n-1)$.
Recalling that the OLS fit corresponds to an orthogonal projection in the $n-$dimensional
space of individual observations onto the column-space of $X$, we can write the following
well-know matrix-expressions:
\begin{equation}
\hat{y} = HH'y = X{\hat{\beta}_0}\text{ .}  \label{OLS}
\end{equation}
\noindent These results follow from writing the singular-value decomposition (SVD)
of $X$ as $X=H\Lambda^{1/2}G'$. $HH'$ then denotes an orthogonal projection; it is a
$n \times n$ symmetric and idempotent matrix of rank $p$ known as the Hat-matrix
for OLS. In particular, $\hat{\beta}_0=Gc$ where $G$ is an orthogonal rotation within the
column-space of $X$, and $c=\Lambda^{-1/2}H'y$ is the $p \times 1$ column vector
containing the \textit{uncorrelated components} of $\hat{\beta}_0$, Obenchain (1975).

Generalized ridge estimators apply scalar-valued \textit{shrinkage-factors}, each confined
to the range $0 \leq \delta_j \leq 1$, to the $p$ uncorrelated components of $\hat{\beta}_0$.
Thus these $\beta-$coefficient estimators are of the form:
\begin{equation}
\text{shrinkage }\hat{\beta} = G{\Delta}c = \sum\limits_{j=1}^p {g_j}{\delta_j}{c_j}\text{ ,}  \label{GRR}
\end{equation}
where $\Delta$ denotes the diagonal matrix containing all $p$ shrinkage $\delta-$factors
and $g_j$ denotes the $j^{th}$ column of $G$.

While the above conventions have placed all $X-$information about the form and extent of any
\textit{ill-conditioning} into a convenient canonical-form, these conventions have
done nothing to predetermine the \textit{relative importance} of individual $x-$variables in
\textit{predicting} $y-$outcomes. That information, as well as information on the
many effects of deliberate shrinkage, may well be best and most-clearly revealed via
\textit{visual} examination of \textit{TRACE diagnostic} plots. For examples, see Figures 1
through 5.  

\section{Quantifying Extent of Shrinkage} 

The \textit{multicollinearity allowance}, $m$, measures the ``extent'' of shrinkage
applied in equation (\ref{GRR}): 
\begin{equation}
m=p-\delta_1-\cdots-\delta_p=rank(X)-trace(\Delta),  \label{MCAL}
\end{equation}
where $0 \leq m \leq p$, Obenchain (1977). Besides being the rank of $X$, $p$ is also the
$trace$ of the OLS Hat-matrix. Similarly, $trace(\Delta)$ is like a measure of ``rank''
for the ($p \times p$) $\Delta$ shrinkage-factor matrix in equation (\ref{GRR}).
Thus $m$ is essentially a measure of \textit{inferred rank deficiency} in the given $X-$matrix.
This deficiency is due to ill-conditioning and is best quantified by using the $\delta$ Shrinkage
Factors that Maximize the Normal distribution-theory Likelihood of the resulting GRR estimates.

Use of this $m-$scale for displaying \textit{TRACE diagnostics} also suggests using
the short-hand notation, $\hat{\beta}_m$, to denote individual $\hat{\beta}$
point-estimates in equation (\ref{GRR}). The OLS solution, $\hat{\beta}_0$, occurs at the
beginning, $m=0$, of each shrinkage path. Similarly, $\hat{\beta}_p\equiv{0}$ denotes the
shrinkage terminus at $m=p$.

Since the range of the $m-$index of equation (\ref{MCAL}) is \textbf{finite}, this $m-$scale
is ideal for use as the horizontal axis on all \textit{TRACE} plots. Specificaly, a \textit{TRACE}
can then display the \textit{entire regularization path}. Since Hoerl and Kennard (1970a,b)
restricted attention to shrinkage-factors of the form $\delta_j = \lambda_j / (\lambda_j + k)$
where $k$ is increased from $0$ to $\infty$, their ridge TRACE display with $k$ on the horizontal
axis could only display some finite, initial fraction of their full path.

\section{MSE Optimal Shrinkage} 

When the unknown \textit{true components} of $\beta$ are denoted by $\gamma = G'\beta$,
it follows that the $i^{th}$ uncorrelated component of the $c-$vector in equation
(\ref{GRR}) has mean $\gamma_i$ and variance $\sigma^2/\lambda_i$.

The unknown true minimum MSE risk value for the $j^{th}$ shrinkage $\delta-$factor is then

\begin{equation} 
\delta _j^{MSE} = \frac{\gamma_j^2}{\gamma_j^2+(\sigma^2/\lambda_j)}
= \frac{\lambda_j}{\lambda_j+(\sigma^2/\gamma_j^2)}
= \frac{\varphi_j^2}{\varphi_j^2+1}\text{ ,} \label{DOPT}
\end{equation}

\noindent where $\varphi _j^2=\gamma_j^2\lambda_j/\sigma ^2$, Obenchain(1975),
equation (2.7). An argument of Hoerl and Kennard (1970a, p. 65) is essentially
equivalent to (\ref{DOPT}) on a generalization of their ``ordinary'' ridge path.

The F-ratio for testing $\gamma _j=0$ is $F_j=(n-p-1)\hat{\rho}_j^2/(1-R^2)$, 
where $\hat{\rho}_j$ denotes the observed \textit{principal correlation} between the
centered and rescaled $y-$vector and the $j^{th}$ column of the $H-$matrix in
equation (\ref{OLS}), while $R^2 = \hat{\rho}_1^2 + \hat{\rho}_2^2 + \cdots +%
\hat{\rho}_p^2$ is the familiar coefficient of determination. Since the unknown
non-centrality of $F_j$ is $\varphi _j^2$, the ML estimator of $\varphi _j^2$ is
$n\cdot F_j/(n-p-1)$ under Normal-theory. 

\section{ML Estimation of Uncorrelated Components} 

When no restrictions are placed on the functional form of regularization, one
is free to simply substitute ML estimates for the unknowns in equation (\ref{DOPT})
to identify the estimate most likely to have minimum MSE risk. This ML shrinkage estimate
under Normal-theory is of the \textit{cubic} (clearly \textit{nonlinear}) form
\begin{equation}
\hat{\gamma}_j^{ML}=\frac{n\cdot \hat{\rho}_j^3}{n\cdot \hat{\rho}_j^2%
+(1-R^2)}\cdot \sqrt{\frac{y^{\prime }y}{\lambda_j}}\text{ .}  \label{URML}
\end{equation}
\noindent Thompson (1968) studied this estimator using numerical integration and
showed that it yields: (i) reduced MSE risk when a true $|\gamma_j|$ is small relative
to $\sigma$, (ii) increased risk when $|\gamma_j|$ is larger, but (iii) the same
limiting risk as $|\gamma_j|$ approaches $+\infty$.

Under conditional distribution-theory for linear models, the $\hat{\gamma}_j^{ML}$
estimates of equation (\ref{URML}) are viewed as being given linear functions of $y$
multiplied by $\sqrt{y^{\prime }y/\lambda_j}$. In other words, the Normal-theory
conditional distributions of $\hat{\rho}_j-$estimates are \textit{not} those of correlation
coefficients. After all, the individual columns of the $H-$matrix are considered given here
and are not subject to random variation. 

In the limit as the OLS estimate of $\sigma^2$ decreases to $0$, $R^2$ naturally
increases to $1$ for a ``correct'' linear model. This causes equation (\ref{URML})
to simplify to
$\hat{\gamma}_j^{ML}=\sqrt{y^{\prime }y}\cdot{\Lambda}^{-1/2}\hat{\rho}=c_j$
when $R^2=1$. This is the special case of equation (\ref{GRR}) where
$\delta_j\equiv 1$, and the OLS fit becomes exact. In other words, OLS predictions,
$X\hat{\beta}_0=\hat{y}$, from equation (\ref{OLS}) are then identical to the observed
$y-$outcomes.

\section{Efficient ML Shrinkage} 

As argued in Obenchain (1975), equation (2.8), the Normal-theory likelihood that
a set of delta shrinkage factors ($\delta_1, \dots , \delta_p$) has minimum MSE risk is
identical with the likelihood that

\begin{equation}
\gamma_j = \pm \hat{\sigma}\{ \delta_j/[\lambda_j(1-\delta_j)]\} ^{1/2}  \label{MLEQN}
\end{equation}
\noindent once this likelihood has been maximized by choice of the $\hat{\sigma}$ estimate of
$\sigma$ as well as by choice of all $p$ of the $\pm$ signs.

In practical applications, equation (\ref{MLEQN}) can be ``too general''. For example, Obenchain
(1975), section 4.1, restricted interest to a $2-$\textit{parameter} family of shrinkage Paths that
frequently fail to pass directly through the \textbf{overall optimal} set of shrinkage
$\delta-$factors of equation (\ref{DOPT}) when $p > 2$.

\begin{figure}
\center{\includegraphics[width=6in]{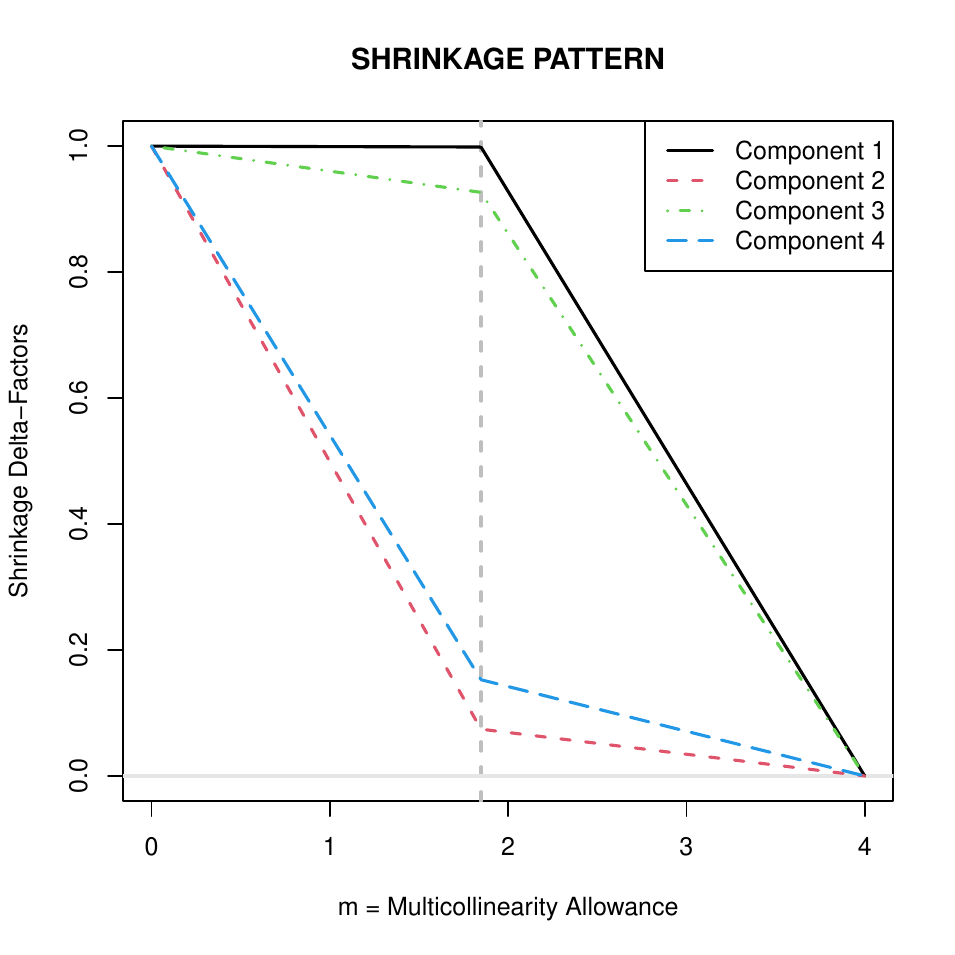}}
\caption{\label{fig:SPAT} Efficient Shrinkage $\delta-$factors for the Portland cement example.
Note that this \textit{TRACE} consists of $p = 4$ two-piece linear splines that all have
their ``interior Knot'' at $m = 1.848$. These $\delta-$factors are applied to the uncorrelated
components vector, $c$, via equation (\ref{GRR}) rather than directly to the OLS $\hat{\beta}-$coefficients
of equation (\ref{OLS}).}
\end{figure}

The new $p-$\textit{parameter} ``efficient'' shrinkage path satisfying equation (\ref{GRR}) and passing
through the \textit{unrestricted} $\hat{\beta}^{ML}=G\hat{\gamma}^{ML}$ estimate defined by equation
(\ref{URML}) is implemented by the \textit{eff.ridge()} R-function, Obenchain (2023). Note that the ``spat''
\textit{TRACE} plot of Figure \ref{fig:SPAT} for this new Maximum Likelihood Path shows exactly why this
Path is ``as short as possible''. After all, the shortest distance between 2 points is a straight line,
so the shortest Path consists of a two-piece linear spline. This Path always starts with the OLS estimate
at $m=0$ and always ends with $\hat{\beta} \equiv 0$ at $m=p$. In Figure \ref{fig:SPAT}, $p=4$, and the
``interior'' knot defined by equation (\ref{URML}) occurs at $m=1.848$.

Specifically, the Efficient Path starts with the OLS solution, $\hat{\beta}_0$, where $m=0$ and all
$\delta-$factors = 1. Similarly, this Path ends with $\hat{\beta} \equiv 0$ at $m = p$ because
each $\delta-$factor is $0$ there. The most-likely shrunken $\beta-$estimates occur at the
interior ``Knot'' that is clearly seen on the ``spat'' Trace. This solution on the efficient-path
cannot be less likely under Normal-theory than the ``best'' solution on any other path. The shapes
of traditional $1-$ or $2-$parameter paths are actually predetermined almost exclusively by the
eigen-values of the $X-$matrix, a disadvantage pointed out in Hoerl and Kennard (1975).

Since Figure \ref{fig:SPAT} depicts the shrinkage $\delta^{\star}-$factors that apply to all
$\hat{\beta}-$estimates for $0 \leq m \leq 4$, note that the four $\delta^{MSE}$ estimates
are $(0.9986, 0.0743, 0.9266, 0.1528)$ here. The second shrinkage-factor, $\delta^{\star}_2$
(red short-dash line), is always smallest because it has the smallest $\delta^{MSE}$ factor.
Similarly, $\delta^{\star}_4$ (blue long-dash line) is also small because $\delta_4^{MSE}$ is
the $2^{nd}-$smallest factor. Ultimately, the $\delta^{\star}_3$ (green dot-dash line) remains
quite close to $\delta^{\star}_1$ (black solid line) because their $\delta^{MSE}-$estimates are
nearly equal and much larger than the other two.

\section{\textit{TRACE} Diagnostics} 

\textit{TRACEs} are graphical aids that help users literally ``see'' most of the details needed
to fully appreciate how and when shrinkage-estimators alleviate the effects of ill-conditioning
on linear model coefficient estimates. Here, we will use a well-known benchmark dataset to
perform calculations and plot \textit{TRACEs}.

The Portland cement dataset of Woods, Steinour and Starke (1932) contains characteristics
of $n=13$ cement mixtures, where the $y-$outcome variable is \textbf{heat} (cals/gm) evolved
during hardening. The $p=4$ predictor $x-$variables recorded are ``ingredient percentages''
that appear to have been ``rounded down'' to full integers. Due to the small size ($n=13$) and
limited number of digits reported, this dataset has served as both a \textit{benchmark} for accuracy
of manual OLS computations and as an example where the ``sign'' of a fitted OLS coefficient
differs from that of the correlation between $y$ and the corresponding $x-$variable.

The sum of all $n=13$ values for all four $x-$predictors varies from 95\% to 99\%. If these
$x-$values had summed to exactly 100\% for all 13 mixtures, the centered X-matrix would then
be of $rank=3$. In other words, this $p = 4$ regression model is rather clearly
\textit{ill-conditioned} in the sense of suffering an effective \textit{rank deficiency} of at
least $m = 1$. In fact, we will see that an $m-$extent of slightly less than $2$ can be a more
appropriate and realistic estimate of \textit{rank deficiency} here.

The \textit{TRACE} diagnostics displayed here in Figures 1 to 5 were generated using the
plot() function for \textit{eff.ridge()} objects in version $2.3$ of the \textit{RXshrink} package
using $steps = 100$. Since calculations defining regularization paths are performed only
on a \textit{lattice} of $m-$extents from equation (\ref{MCAL}), each $m-$value is then
a multiple of $1/100=0.01$. The default setting for \textit{eff.ridge()} uses
$steps = 8$ to produce a unit change in $m$. The extra ``detail'' that results from
$steps = 100$ shows that the curved portions of Figures 3, 4 and 5 are very smooth.

\begin{figure}
\center{\includegraphics[width=6in]{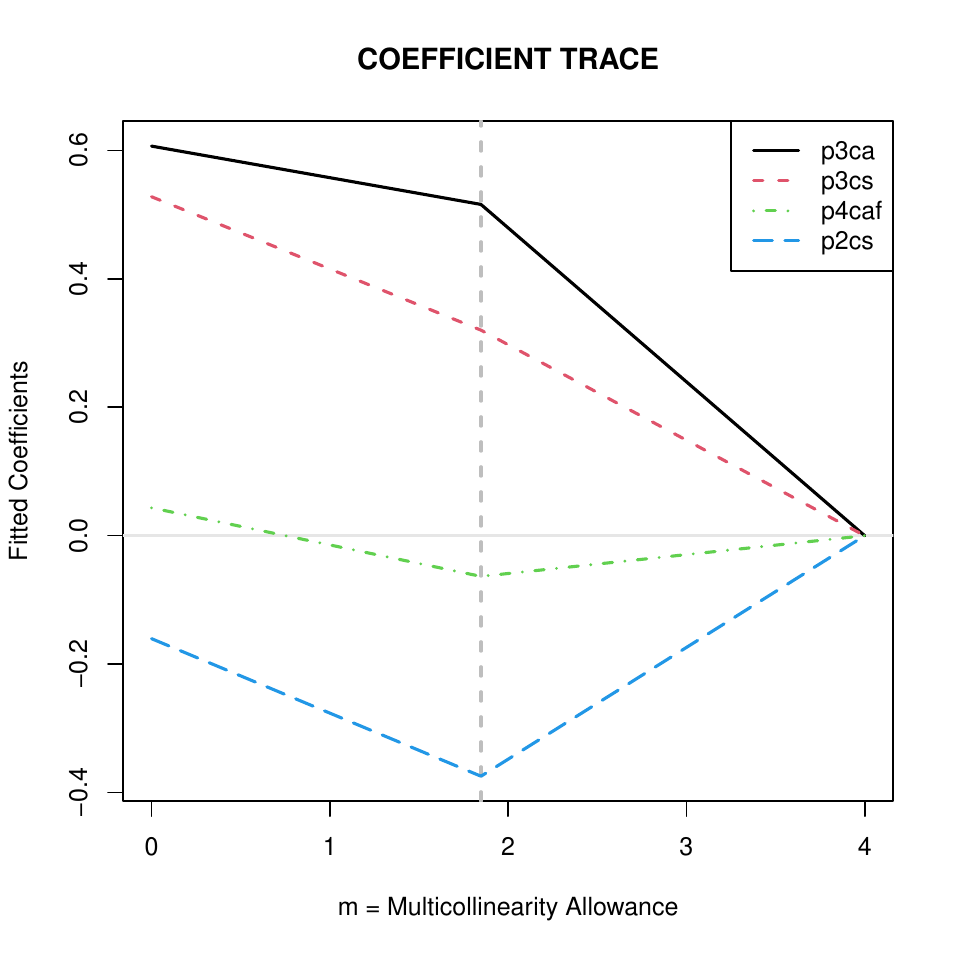}}
\caption{\label{fig:COEF} Efficiently shrunken $\beta-$coefficient estimates for the Portland
cement data. This \textit{TRACE} also consists of $p = 4$ two-piece linear splines. The vertical
dashed-line again marks the ``Knot'' at $m = 1.848$. Note the ``numerical sign'' correction
to the green dot-dash ``p4caf'' coefficient on this TRACE when the $m-$extent of shrinkage
exceeds $m = 0.76$. The $LP = 7$ (Legend Position) here is ``Upper-Right''.}
\end{figure}

For the Portland cement data, the ML shrinkage-extent is $m = 1.848$. Thus a gray vertical dashed
line is plotted at this point on the \textit{TRACEs} displayed in Figures 1 to 5. 

Note that Figure \ref{fig:COEF} also features a ``wrong sign'' \textit{correction} to the
$3^{rd}$ fitted coefficient (percentage of \textbf{p4caf} in the mix, green dot-dashed line). This
$3^{rd}$ coefficient becomes negative past $m=0.75$ to agree in sign with the marginal correlation
($-0.5347$) of \textbf{p4caf} with the \textbf{heat} $y-$outcome.

Note that he \textit{numerical signs} and \textit{relative magnitudes} of $\hat{\beta}_m$ estimates in
Figure \ref{fig:COEF} change between $m=0$ and the ``Interior Knot'' at $m=1.848$ on the
Efficient Path. However, from that point onward to the Shrinkage Terminus at $m=p$, note that both the
\textit{numerical signs} and \textit{relative magnitudes} of shrunken coefficients in their TRACE
displays \textbf{always remain perfectly stable}.

All $p-$curves plotted in the efficient \textit{Coefficient TRACE} of Figure \ref{fig:COEF}, like
the corresponding \textit{Shrinkage Pattern TRACE} of Figure \ref{fig:SPAT}, are always
\textbf{two-piece linear splines}. In sharp contrast, the three other types of \textit{TRACE} plots
need to contain ``curved'' lines to realistically depict the \textit{Non-Linear Effects} of
shrinkage on measures of \textit{MSE Risk} and Normal-theory \textit{Likelihood Ratio} statistics.

\begin{figure}
\center{\includegraphics[width=6in]{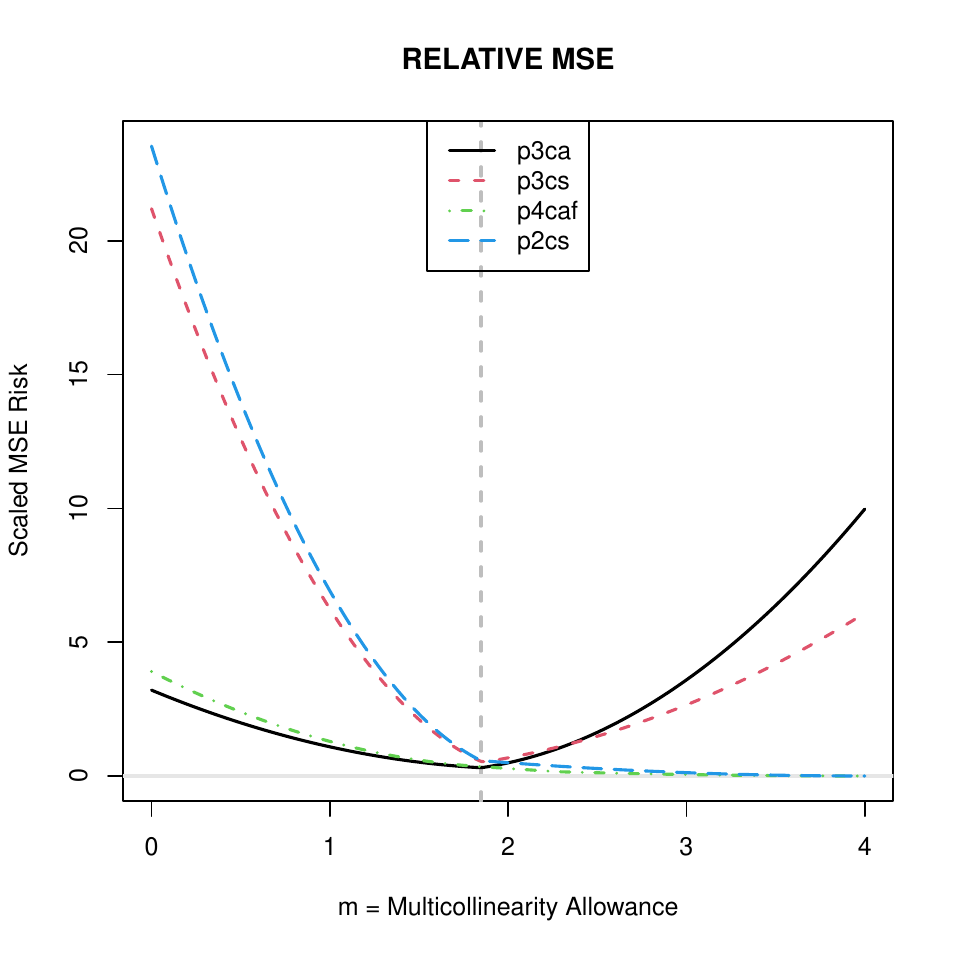}}
\caption{\label{fig:RMSE} Unrestricted relative MSE risk for the Portland cement
data. The curves on this \textit{TRACE} look somewhat quadratic because the unknown
true relative-risk curves are exactly quadratic! The $LP = 6$ (Legend Position) here
is ``Upper-Center''.}
\end{figure}

\begin{figure}
\center{\includegraphics[width=6in]{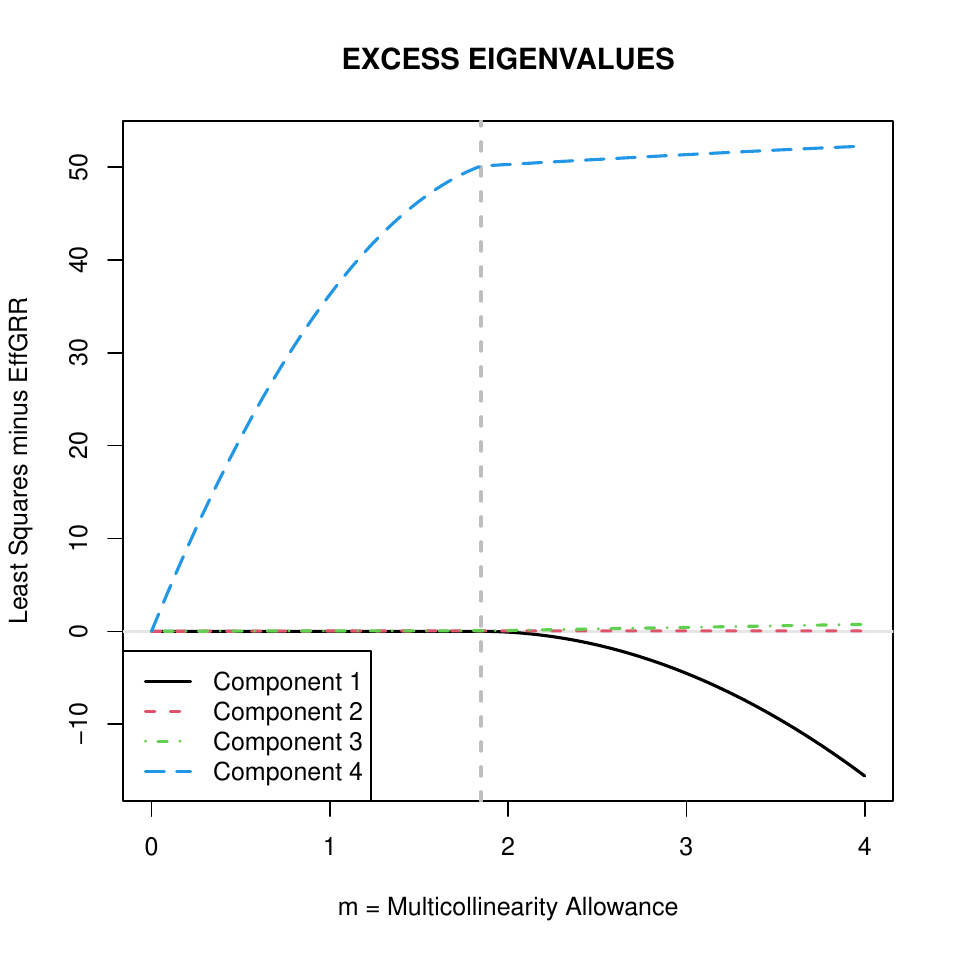}}
\caption{\label{fig:EXEV} Unrestricted Excess Eigenvalues for the Portland cement
data. A positive Excess Eigenvalue emerges as shrikage occurs to the right of $m=0.0$
and grows rapidly until shrinkage exceeds the MSE risk optimal extent, $m=1.848$. The
bad-news is that the single possible \textit{negative} Excess Eigenvalue appears for
$m > 1.848$ and slowly becomes more and more negative as shrinkage continues. The
$LP = 3$ (Legend Position) here is ``Lower-Left''.}
\end{figure}

\begin{figure}
\center{\includegraphics[width=6in]{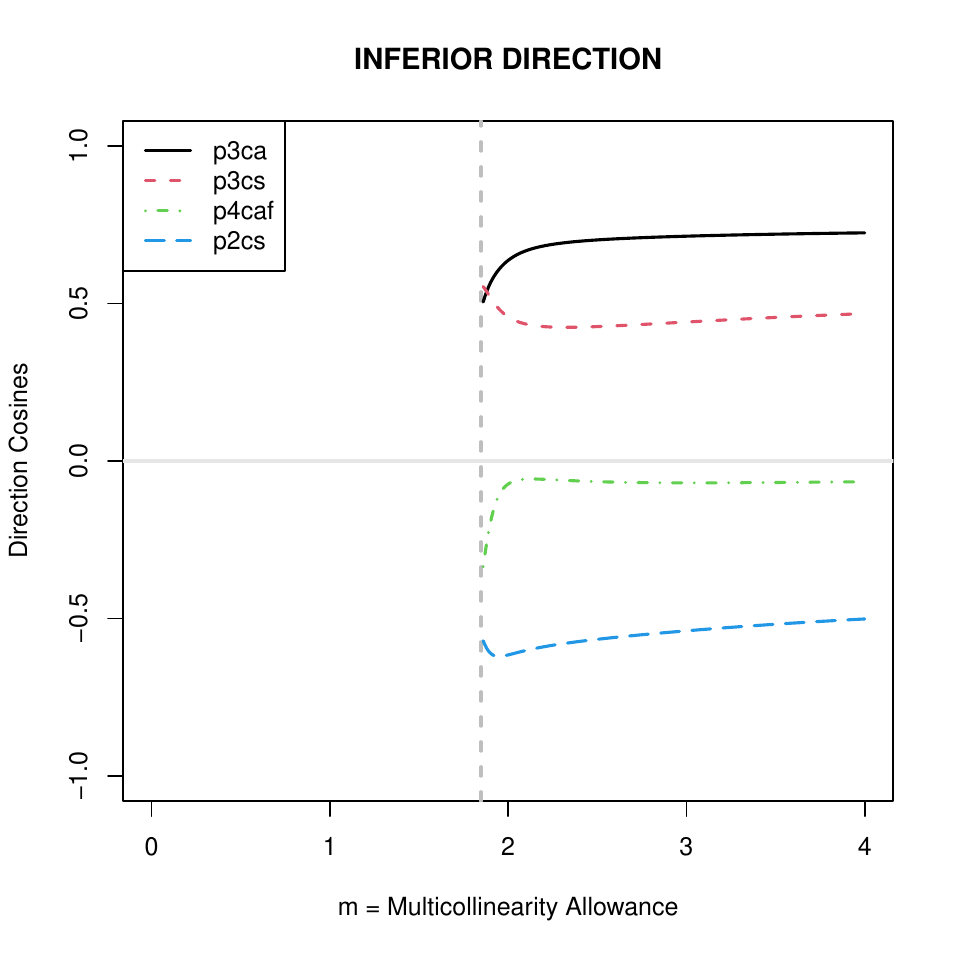}}
\caption{\label{fig:INFD} Unrestricted Inferior Direction-cosines for the Portland
cement data. Note that there is no ``Inferior Direction'' to the left of $m=1.848$.
However, one does suddenly appear to the right of $m=1.848$, and it then
rotates slightly in $4-$dimensional $x-$space until it reaches its terminal orientation
at $m=4$. The $LP = 5$ (Legend Position) here is ``Upper-Left''.}
\end{figure}

The \textit{relative MSE TRACE} of Figure \ref{fig:RMSE}, the
\textit{excess eigenvalue TRACE} of Figure \ref{fig:EXEV} and the
\textit{inferior direction TRACE} of Figure \ref{fig:INFD} are all
based upon risk-related ML estimators.
In particular, relative-risk estimates are given by the diagonal elements of the
$MSE/\hat{\sigma}^2$ matrix and are both particularly relevant and easy to
interpret in Figure \ref{fig:RMSE}. While unbiased under Normal-theory, each
estimated relative-risk is increased, if necessary, to assure it is at least as large
as its relative-variance, $\hat{\delta_j^2}/\lambda_j$. Here, user
interest rightly becomes focused upon the range $1.5{\leq}m{\leq}2.5$ where
relative-risks are greatly reduced.

The eigen-values and vectors of the difference in risk matrices, \{$MSE(ols)-MSE(ridge)$\},
provide clear insights into key effects of ridge shrinkage. The good news is that
at most \textit{one eigenvalue} of this difference in MSE risks can be \textit{negative},
Obenchain (1978). While an ``inferior direction'' corresponding to a negative
estimated excess-eigenvalue does suddenly appear in Figure \ref{fig:INFD} at
$m{\approx}1.8$, the largest \textit{positive} excess eigenvalue in
Figure \ref{fig:EXEV}  is relatively gigantic ($+50$) at this same $m-$extent.
The only negative excess-eigenvalue indicates a MSE risk \textit{increase}
due to shrinkage of at most $15.6$ even at $m=4$, while the concomitant
\textit{decrease} in MSE in a direction strictly orthogonal to this lone inferior
direction exceeds $+50$. Thus, shrinkage along the path depicted in Figure
\ref{fig:SPAT} has clear potential for a \textit{net overall reduction} in MSE risk.

Also note that the inferior-direction at the shrinkage terminus, $m=4$, of
Figure \ref{fig:INFD} ends up pointing almost directly ``backwards'' at the initial OLS
$\hat{\beta}_0$ solution of Figure \ref{fig:COEF}. The absolute value of the
correlation between the two corresponding direction-cosine vectors is $0.988$ here,
and similar results would always be expected whenever the initial OLS solution
is significantly different from the shrinkage terminus, $\hat{\beta}_p\equiv{0}$,
at $m=p=4$.

\begin{figure}
\center{\includegraphics[width=6in]{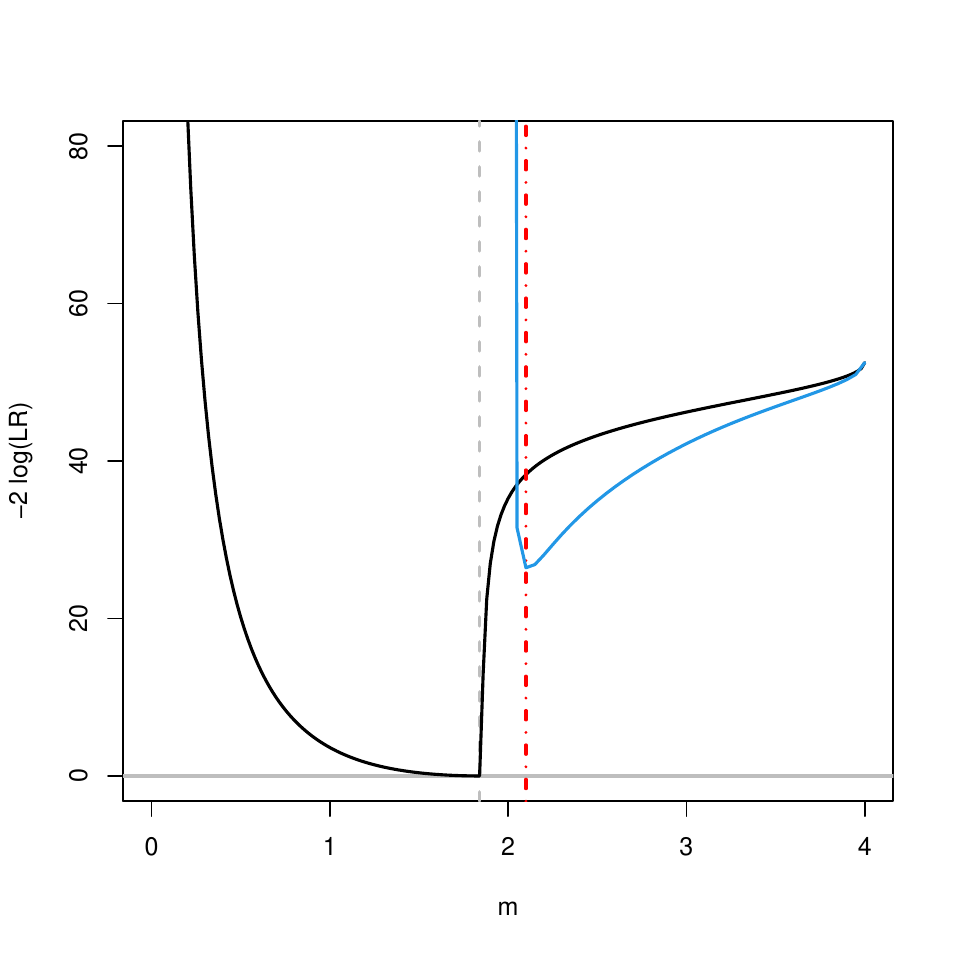}}
\caption{\label{fig:LRAT} Negative Log Likelihood Ratio curves for two different
shrinkage-paths on the Portland cement data. The solid black curve shows how the
$-2$ log Likelihood-Ratio under Normal-theory drops all of the way to $0$ at $m=1.848$
for the ``Efficient'' $4-$parameter path. The solid blue curve depicts
the corresponding LR for the $2-$parameter path of ``best'' $q-\text{Shape}=-5$.
This $-2 log(LR)$ reaches a minimum of $26.4$ at $m\approx{2.1}$. Since the upper $99\%-$point
of a $\chi^2-$variate with $2$ degrees-of-freedom is only $9.21$, this ``best''
$2-$parameter fit is significantly different from MSE-optimal.}
\end{figure}

Finally, Figure \ref{fig:LRAT} displays a type of plot not available within the
\textit{RXshrink} R-package because it compares results for 2 different shrinkage paths
for the Portland cement data. The horizontal axis is $m-$Extent of Shrinkage, while the
vertical axis is $-2 log$(Likelihood Ratio) over the finite range [$0, 80$]. 
The \textit{eff.ridge()} and \textit{qm.ridge()} R-functions were invoked with ``steps=100''.
In Figure \ref{fig:LRAT}, the solid-black curve shows how the Likelihood Ratio $\chi^2$
for the efficient path swoops down from very large values all of the way to $0.0$ at
$m=1.848$, then suddenly starts increasing and ultimately ends at $52.5$ for $m=4$. The
solid blue curve that enters the top of Figure \ref{fig:LRAT} at $m \approx 2$ corresponds
to the Likelihood Ratio $\chi^2$ for the \textit{qm.ridge()} path of most-likely
$q-\text{Shape}=-5$ on the default mesh. This solid blue curve reaches its minimum of $26.4$
at $m=2.1$ [marked by a vertical red dot-dash line], then also increases to $52.5$ at $m=4$.
Figure \ref{fig:LRAT} thus confirms that the \textit{eff.ridge()} path is fully
\textbf{efficient}. After all, this path reduces the Normal-theory Negative Log Likelihood
Ratio to \textbf{Zero} while using an equal or lesser $m-$Extent of shrinkage than all viable
alternatives.

Truly ``favorable'' cases of ridge shrinkage occur when an $m-$Extent greater
than 1.0 is favored in \textbf{two senses:} (a) no ``inferior direction'' has yet
appeared, and (b) the relative MSE risk over all $p$ coefficients is optimized. Both
indicators are clearly present in the \textit{TRACE} plots of Figures \ref{fig:EXEV},
\ref{fig:INFD} and \ref{fig:RMSE} for the Portland cement data benchmark.

\section{Elliptical Confidence Regions for Pairs of Coefficients} 

One of the central ideas in generalized ridge regression is that unbiasedness is not
a very important property in \textbf{point estimation} of regression $\beta-$coefficients.
As shown in Figure \ref{fig:RMSE}, biased point estimates can have smaller Mean Squared
Error risk than unbiased OLS estimators when shrinkage reduces variance by more than the
concomident increase in squared bias.

On the other hand, unbiasedness plays a more central role in the Classical theory of
\textbf{set estimation} (statistical inference) than it does in point estimation. For example,
an unbiased set estimator will not cover any incorrect value with larger probability than
the probability, $1-\alpha$, of covering the correct value; biased set estimators will
cover some incorrect values with probabilities greater than $1-\alpha$. See Obenchain (1977).

\begin{figure}
\center{\includegraphics[width=6in]{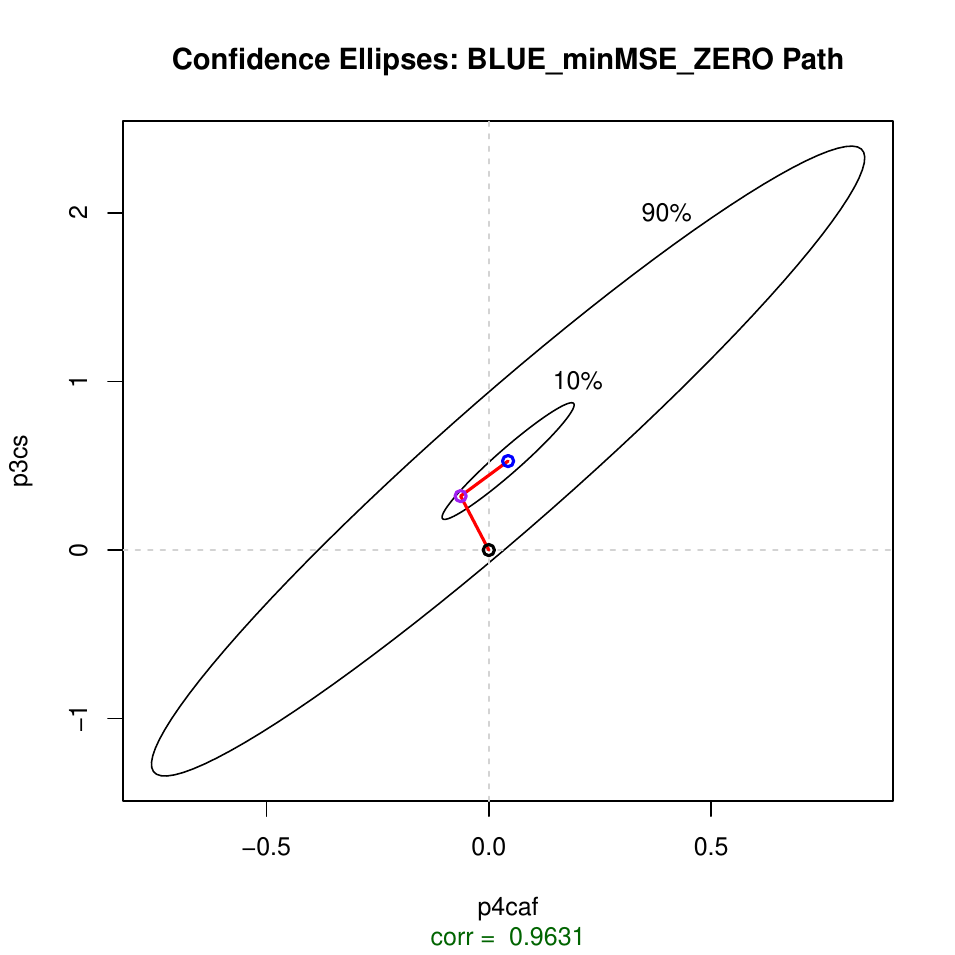}}
\caption{\label{fig:BIVCR} Here are two unbiased Elliptical Confidence regions for the
Second (``p3cs'') and Third (``p4caf'') regression coefficients in the $p = 4$ linear model with
TRACEs depicted in Figures 1 to 5. Note that the change in sign of the ``p4caf'' coefficient
occurs within the inner Ellipse of only $10\%$ confidence. Furthermore, since the point where
the ``p3cs'' and ``p4caf'' estimates are both Zero lies within the $90\%$ confidence Ellipse,
at least one of these two coefficients is not significantly different from Zero with confidence of 
$10\%$ or less. Here, the ``p4caf'' coefficient that changes sign is \textit{not} significant,
while the ``p3cs'' coefficient is highly significant (***). See ANOVA Table $1$.} 
\end{figure}

Figure \ref{fig:BIVCR} was created using the \textit{eff.aug()} and \textit{eff.biv()} functions
in version $2.3$ of the \textit{RXshrink} R-package to illustrate graphics useful in Classical
statistical inference on the effects of shrinkage. Purely Bayesian inferences are quite different
because they view shrinkage estimators as being unbiased relative to \textit{combined Sample and
Prior information}.

\begin{table}[h!]
\centering
\begin{tabular}{@{}llllllr@{}} 
\multicolumn{7}{c}{TABLE 1 -- ANOVA Table} \\
Variable & DF & Sum Sq. & Mean Sq. & F Value  & Pr($\textgreater$F) & \\
p3ca     &  1 & 1450.08 & 1450.08  & 242.3679 & 2.888e-07 & *** \\
p3cs     &  1 & 1207.78 & 1207.78  & 201.8705 & 5.863e-07 & *** \\   
p4caf    &  1 &    9.79 &    9.79  &  1.6370  & 0.2366    &     \\       
p2cs     &  1 &    0.25 &    0.25  &  0.0413  & 0.8441    &     \\
\end{tabular}
\end{table}

\section{Linear Models of Rank One or Two} 

\begin{figure}
\center{\includegraphics[width=6in]{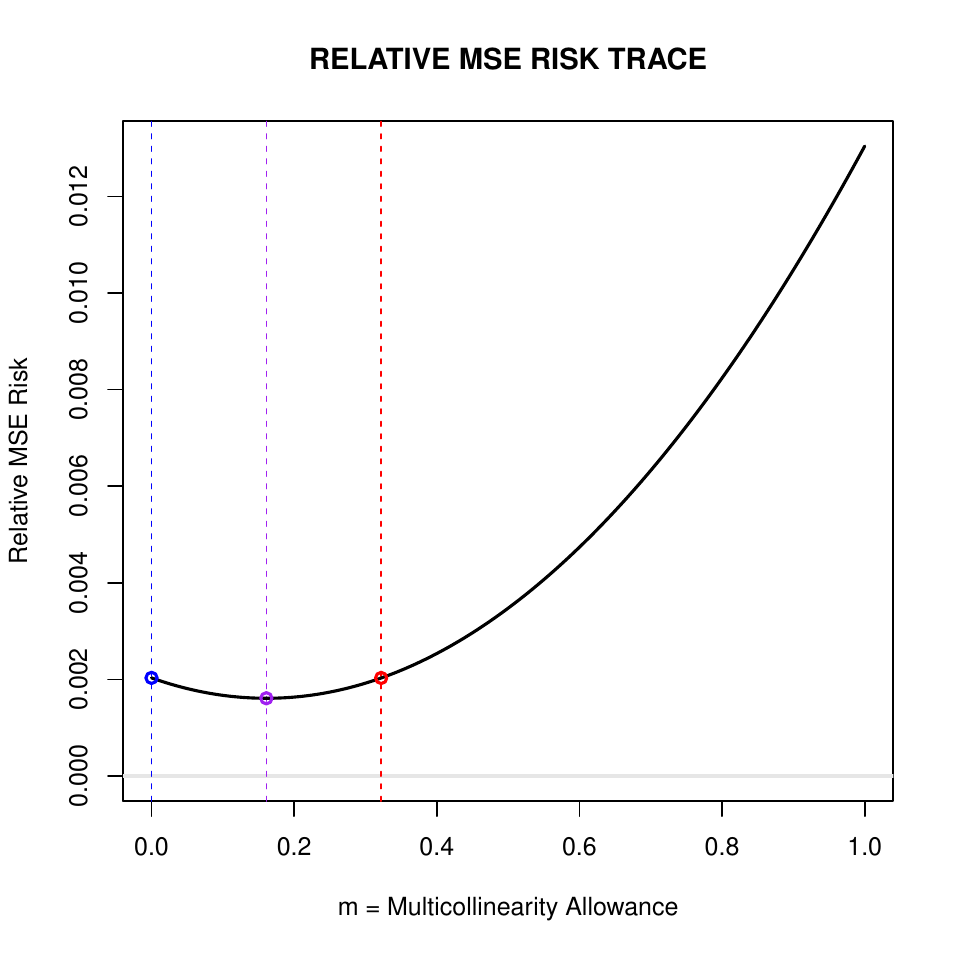}}
\caption{\label{fig:YonXrisk} The ``rmse'' TRACE for the ``Simple'' Linear Regression of \textit{heat}
onto \textit{p4caf} shows a single quadratic curve with its minimum at $m = 1 - \delta^{MSE} = 0.161$.
Furthermore, when the m-Extent of shrinkage is doubled to $m = 0.322$, note that the resulting biased
$\beta-$estimate has the same Relative MSE Risk as the unbiased OLS solution at $m = 0$, which is
also the Relative Variance of OLS.} 
\end{figure}

Shrinkage of ``Simple'' ($p = 1$) Linear Regression models (with R-code formula $y \sim x$),
where $x$ is a non-constant vector, are performed by the \textit{YonX()} R-function,
Obenchain (2021). All statistics that are ($p \times 1$) vectors or ($p \times p$) matrices when
$p > 1$, are simply scalars in these $p = 1$ models, and the relative MSE risk is strictly quadratic
in these cases. As a result, when the \textit{heat} y-Outcome is regressed on only the percentage of
\textit{p4caf} in Portland cement mixtures, the unbiased OLS estimate of the $\beta-$coefficient,
$-1.256$ cannot have the ``wrong'' numerical sign, and the ``rmse'' TRACE of Figure \ref{fig:YonXrisk}
is purely quadratic. Any MSE risk ``inferior-direction'' when $p=1$ is simply ``up and down.''  

\begin{figure}
\center{\includegraphics[width=6in]{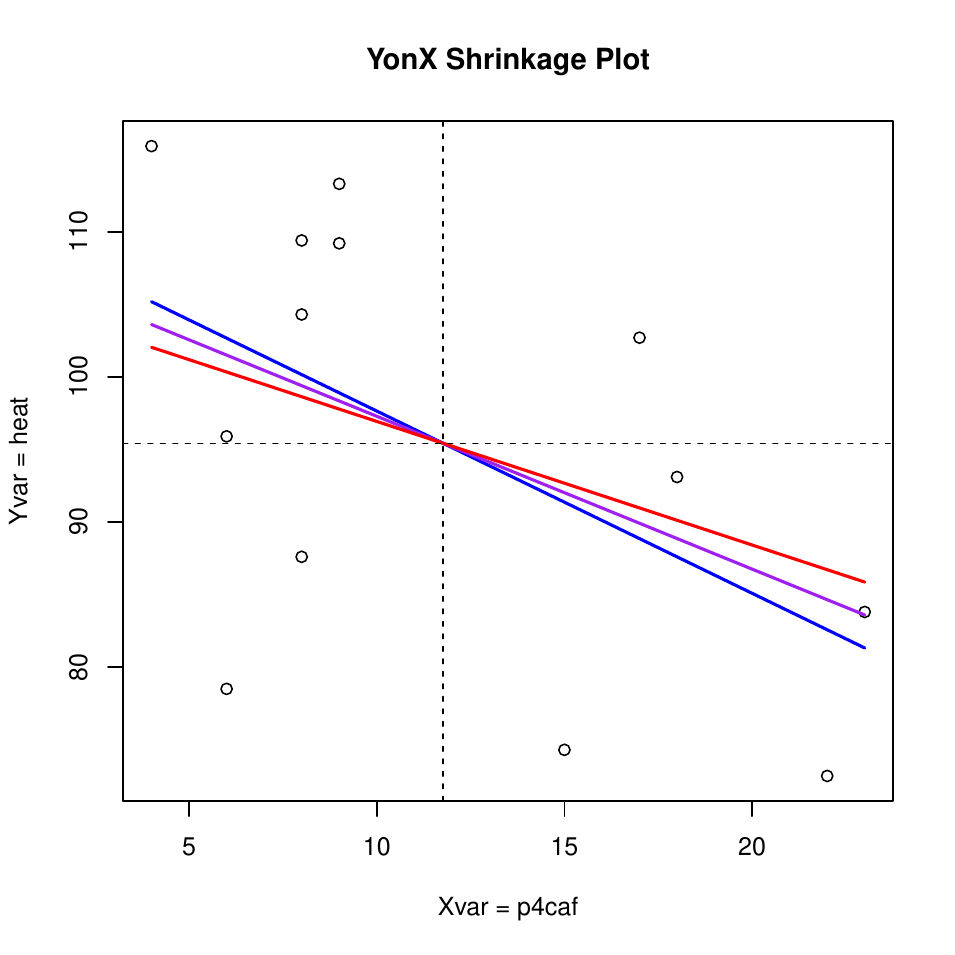}}
\caption{\label{fig:YonXfit} Three different slopes for the ``Simple'' Linear Regression model
(heat $\sim$ p4caf) are shown here superimposed upon the scatter of 13 Portland cement mixtures. Note
that ``shrinkage'' always tends to make the Purple and Red fitted lines more horizontal than the Blue fit.} 
\end{figure}

TRACE plots for $p=1$ models are also augmented with a graphic like that of Figure \ref{fig:YonXfit} that
shows the observed (x=p4caf, y=heat)$-$scatter plus three fitted lines that pass through ($\bar{x}, \bar{y}$).
The OLS fitted line is colored Blue [Best Linear Unbiased], the Minimum MSE Risk line ($m=0.161$) is colored
Purple, and the ``Double Shrunken'' line ($m=0.322$) is colored Red.

Models with $p=2$ non-constant X-variables, use generalized ridge shrinkage $\delta-$factors
that can be written in the form $\delta_i =1/(1+f_i)$ where $f_1$ and $f_2$ are both
\textit{non-negative}. For example, $q-\text{Shape}$ paths, Goldstein and Smith (1974),
are of this form and can pass through the ML coefficient-vector with MSE optimal risk.
This requires

\begin{equation}
f_i = k \lambda_i ^{q-1} = \sigma ^2 \lambda_i ^{-1} /\gamma_i ^{2}\text{ ,} \label{eqn:fMSE}
\end{equation}

\noindent for $i = 1$ and $2$, where the $\gamma_i$ values are unknown but can easily be
estimated. Note that the k-factor is a non-negative constant that starts at $k=0$
and is steadily increased to produce shrinkage.

The restrictions imposed via equation (\ref{eqn:fMSE}) suggest using a $q-$Shape of the form:

\begin{equation}
\hat{q}^{MSE} = - \frac{ln(\hat{\gamma}_1 ^2 / \hat{\gamma}_2 ^2)}{ln(\lambda_1 ^2 / \lambda_2 ^2)} \text{ .} \label{qGSeqn}
\end{equation}

Two invocations of the \textit{qm.ridge()} R-function can be used to perform ``best'' MSE
risk computations and display Traces for rank $p=2$ models. An initial call of the form
\textit{qm.ridge($y \sim x1 + x2$, data)} uses the default argument of $Q =$ ``qmse'' to make a
simple mesh search for a reasonable $q-$Shape. But, when $p = 2$, the printed output also
displays an estimate, $Qbest$, of $q^{MSE}$ from equation (\ref{qGSeqn}). A second call of
the form \textit{qm.ridge($y \sim x1 + x2$, data, $Q = Qbest$)} then uses this ``best''
\textit{smoothly curved path}. Naturally, this path is slightly longer than the efficient
path from the call: \textit{eff.ridge($y \sim x1 + x2$, data)}. For example, the Optimal
Q$-$Shape for the model ``\textit{heat} $\sim$ \textit{p3cs + p2cs}'' on the Portland cement
data.frame is Q$=-0.6953$.

\section{Updates and Corrections to Earlier Versions} 

A vast majority of the material included here (plus a $\S 10$ on ``Bi-variate Distributions of
Non-Linear Beta-Coefficient Estimates'') was published in \textbf{Open Statistics}, Obenchain
(2022). Since that journal is not currently being published, this latest revision ($v5$) of my
2021 arXiv paper provides some updates as well as a \textbf{correction} to the caption of Figure 7.

The default for the $plot(eff.ridge, trace, LP)$ directive in version $2.3$ of the \textit{RXshrink}
$R-$package, Obenchain (2023), is $LP = 0$ to display a TRACE without any ``Legend''. As shown below
in Table $2$, $9$ positive integer values for the ``LP'' argument are recognized. 

\begin{table}[h!]
\centering
\begin{tabular}{@{}lllr@{}} 
\multicolumn{4}{c}{TABLE 2 -- Legend Positions Table} \\
        & Left & Middle & Right \\ 
Upper   & $LP = 5$ & $LP = 6$ & $LP = 7$ \\
Middle  & $LP = 4$ & $LP = 9$ & $LP = 8$ \\   
Bottom  & $LP = 3$ & $LP = 2$ & $LP = 1$ \\       
\end{tabular}
\end{table} 

\section{Summary} 

When linear models are fit to ill-conditioned or confounded \textit{narrow-data},
\textit{TRACE} plots are useful in demonstrating and justifying deliberately
\textit{biased} estimation. This makes plots of \textit{TRACE diagnostics} powerful
``visual'' displays. If advanced students of regression are trained in TRACE interpretation,
they can help administrators capable of basic statistical thinking to avoid misinterpretations
of questionable regression coefficient estimates.

All five types of ridge \textit{TRACE} plots for a wide variety of ridge \textit{paths}
can be explored using existing R-functions. For example, the RXshrink \textit{aug.lars()} function
generates \textit{TRACE}s for Least-Angle, Lasso and Forward Stagewise methods [Efron, Hastie,
Johnstone and Tibshirani (2004), Hastie and Efron (2013)] when applied to \textit{narrow-data}
with $p<(n-3)$. These \textit{TRACE} plots provide quick and clear insights into the MSE
risk characteristics of shrinkage and/or selection methods.

Computers have shaped the \textit{theory} as well as the \textit{practice} of
statistics ever since Efron (1979) helped initiate the emergence of data-science.
Software providing clear ``visual insights'' into the strengths and weaknesses of
alternative estimation methods are indispensable components of an adequate \textbf{Tool Bag}
for tomorrow's applied researchers and data-scientists.

\section{References}

Efron, B. (1979). ``Computers and the Theory of Statistics: Thinking the Unthinkable.''
\textit{SIAM Rev} 21, 460$-$480.\\

\noindent Efron, B., Hastie, T., Johnstone, I., and Tibshirani, R. (2004). ``Least Angle Regression.''
\textit{Annals of Statistics} 32, 407$–$499. (including discussion)\\

\noindent Gunst, R. F. (2000). ``Classical Studies That Revolutionized the
Practice of Regression Analysis.'' \textit{Technometrics} 42, 62$-$64.\\

\noindent Hastie, T., and Efron, B. (2013). ``\textit{lars}: Least Angle Regression,
Lasso and Forward Stagewise.'', ver 1.2, https://CRAN.R-project.org/package=lars\\

\noindent Hoerl, A. E. and Kennard, R. W. (1970a). ``Ridge Regression:
Biased Estimation for Nonorthogonal Problems.'' \textit{Technometrics} 12,
55$-$67.\\

\noindent Hoerl, A. E. and Kennard, R. W. (1970b). ``Ridge Regression:
Applications to Nonorthogonal Problems.'' \textit{Technometrics} 12,
69$-$82.\\

\noindent Hoerl, A. E. and Kennard, R. W. (1975). ``A Note on a Power
Generalization of Ridge Regression.'' \textit{Technometrics} 17, 269.\\

\noindent Obenchain, R. L. (1975). ``Ridge analysis following a
preliminary test of the shrunken hypothesis.'' \textit{Technometrics} 17,
431$-$441. http://doi.org/10.1080/00401706.1975.10489369
(with discussion by G. C. McDonald, 443$-$445.)\\

\noindent Obenchain, R. L. (1977). ``Classical F-tests and confidence regions for ridge regression.''
\textit{Technometrics} 19, 429$-$439. http://doi.org/10.1080/00401706.1977.10489582\\

\noindent Obenchain, R. L. (1978). ``Good and Optimal Ridge Estimators.''
\textit{Annals of Statistics} 6, 1111$-$1121. http://doi.org/10.1214/aos/1176344314\\

\noindent Obenchain, R. L. (2022), ``Efficient Generalized Ridge Regression'', \textit{Open Statistics} 
3(1), 1$-$18. https://www.degruyter.com/document/doi/10.1515/stat-2022-0108/html\\

\noindent Obenchain, R. L. (2023). ``\textit{RXshrink}: Maximum Likelihood Shrinkage using Generalized
Ridge or Least Angle Regression Methods'', ver 2.3, https://CRAN.R-project.org/package=RXshrink\\

\noindent Thompson, J. R. (1968). ``Some shrinkage techniques for estimating the mean.''
\textit{Journal of the American Statistical Association} 63, 113$-$122.\\ 

\noindent Woods, H., Steinour, H. H., and Starke. H. R. (1932). ``Effect of composition of Portland cement
on heat evolved during hardening.'' \textit{Industrial Engineering and Chemistry} 24, 1207$-$1214.\\

\end{document}